\documentclass{appolb}
\usepackage[dvips]{graphicx}
\usepackage{cite}

\begin{document}
\title{SEARCH FOR THE $\eta$-MESIC NUCLEI BY MEANS OF COSY-11, WASA-at-COSY and COSY-TOF 
DETECTOR SYSTEMS \footnote{Presented at the International Symposium on Mesic Nuclei, Kraków, Poland, June 16, 2010.}
}
\author{
P.~Moskal$^{a,b}$\footnote{\texttt{p.moskal@uj.edu.pl}}, J. Smyrski$^{a}$
\address{
$^{a}$ Institute of Physics, Jagiellonian University, Cracow, Poland\\
$^{b}$ Institut f\"ur Kernphysik and J{\"u}lich Center for Hadron Physics,\\
Forschungszentrum J\"ulich, J\"ulich, Germany\\
}
}
\maketitle
\begin{abstract}
We review status and perspectives of the search for the light eta-mesic 
nuclei using COSY-11, WASA-at-COSY and COSY-TOF detector systems.
\end{abstract}
\PACS{13.60.Le; 14.40.Aq}
  
\section{Introduction}
The negatively charged pions and kaons can be trapped 
in the Coulomb potential of atomic nucleus  forming so called pionic (kaonic) atoms.
Observations of such atoms allows for studies of strong interaction of pions and kaons
with atomic nuclei on the basis of shifts and widths of the energy levels~\cite{strauch, zmeskal}.
 
It is also conceivable that a neutral meson could be bound to a nucleus. 
In this case the binding is exclusively due to the strong interaction 
and hence such object can be called a {\it mesic nucleus}. 
Here the most promising candidate
is the $\eta$-mesic nucleus since the $\eta$N interaction 
is strongly attractive.  
As discussed in the contribution by Wycech~\cite{wycechetamesic} 
the nuclear states of $\eta$ may be bound by a  similar mechanism as states of 
$\overline{K}$ with the difference that in the case of $\overline{K}$ 
nucleons are excited to the 
$\Lambda(1405)$
and in the case of $\eta$ meson to the $N^*(1535)$ resonant state.  
We may thus  picture~\cite{sokol0106005} the formation and decay of the
$\eta$-mesic nucleus
as the $\eta$ meson absorption by one of the nucleons 
leading to the creation of the $N^*(1535)$ and 
then its propagation through the nucleus until it decays into 
the pion-nucleon pair 
which escapes from the nucleus.
Predicted values of the width of such states range 
from $\sim$7 to $\sim$40~MeV~\cite{osetPLB550,liu3,haideracta}.
However, it is 
important to stress that the width and binding
energy of the eta-nucleus bound states depends strongly on the not well known
subthreshold  $\eta$-nucleon interaction~\cite{liu3}
making impossible an unambiguous prediction as regards the  bound states,
and, therefore,
direct measurements are necessary~\cite{liu3}.

The search of the $\eta$~-~mesic nucleus was conducted
in many inclusive experiments~\cite{bnl,lampf,lpi,jinr,gsi,gem,mami}
via reactions induced by pions~\cite{bnl,lampf}, 
protons~\cite{jinr,gem}, and photons~\cite{lpi,mami}.
Many  promising indications 
of the existence of such an object
were reported~\cite{lpi,mami,gem},
but so far none was independently confirmed.  
Experimental investigations with high statistical sensitivity
and the detection of the N$^*(1535)$ decay products are being continued at 
the COSY~\cite{wasaproposal,wasaacta}, 
JINR~\cite{jinr}, J-PARC~\cite{fujiokasymposium}, and
MAMI~\cite{kruscheacta} laboratories.
In this contribution we report on searches of the $\eta$~-~mesic helium 
in  exclusive measurements carried out at the cooler 
synchrotron COSY by means of the 
WASA-at-COSY~\cite{wasaproposal,wasaacta}  
and COSY-11~\cite{c11acta,c11meson08,c11nuclphys,c11aps}     
detector setups. We present also suggestion of studies of $\eta$-mesic tritium with quasi-free reactions 
at the COSY-TOF facility.

We consider the study of the $\eta$-mesic nuclei as
interesting on its own account, but  additionally  
it is useful for investigations of
(i) the $\eta$N interaction, 
(ii) the N$^*(1535)$ properties in nuclear matter~\cite{jido},  
(iii) the properties of the $\eta$ meson
in the nuclear medium~\cite{osetNP710}, 
and 
(iv) the flavour singlet component  
of the $\eta$ meson~\cite{bass,basssymposium}. 

\section{Indications for the existence of the $\eta$~-~mesic helium}
\begin{figure}[h] 
    \begin{center} 
        {\includegraphics[scale=0.300]{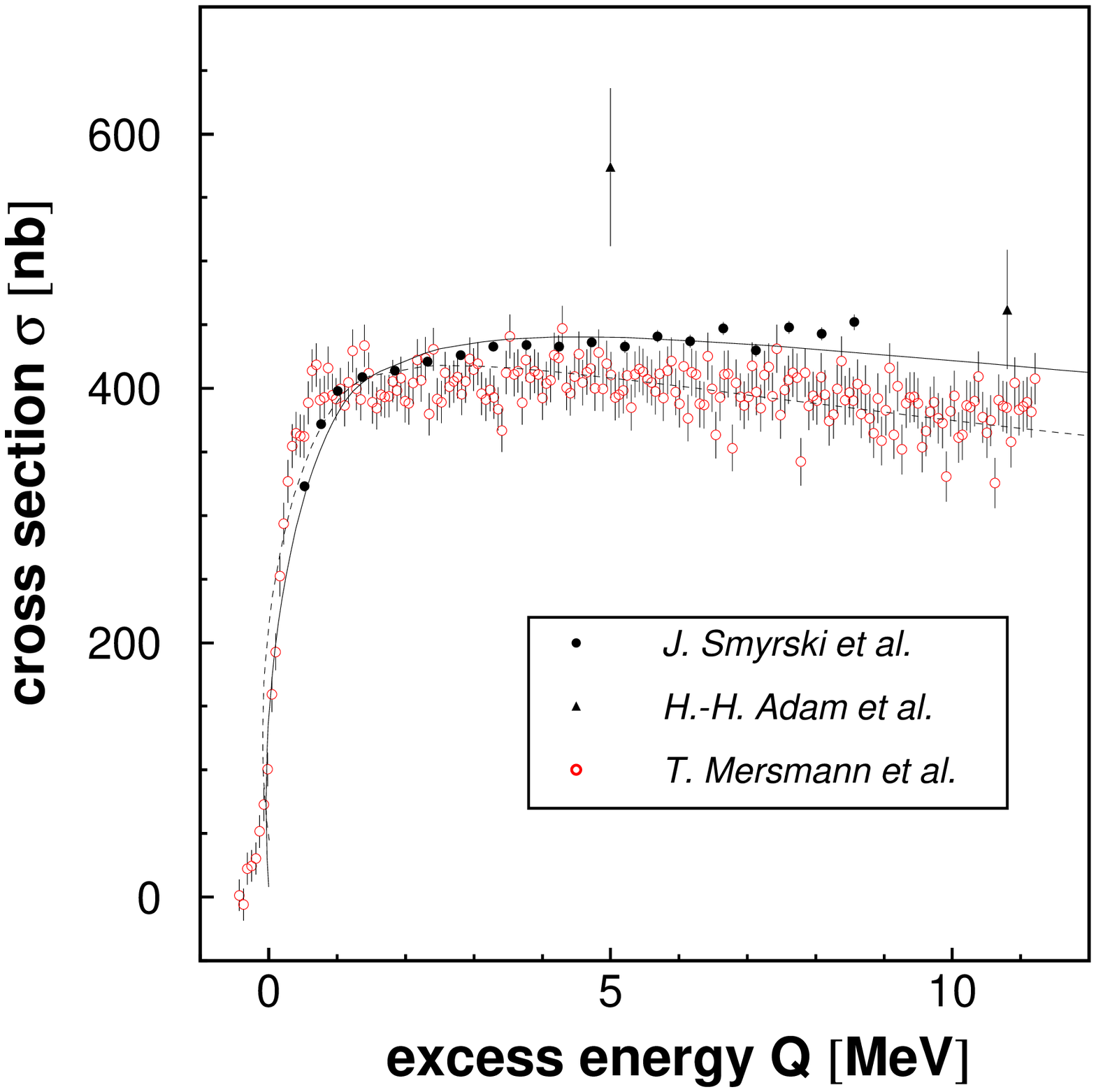}} \hfill
        {\includegraphics[scale=0.303]{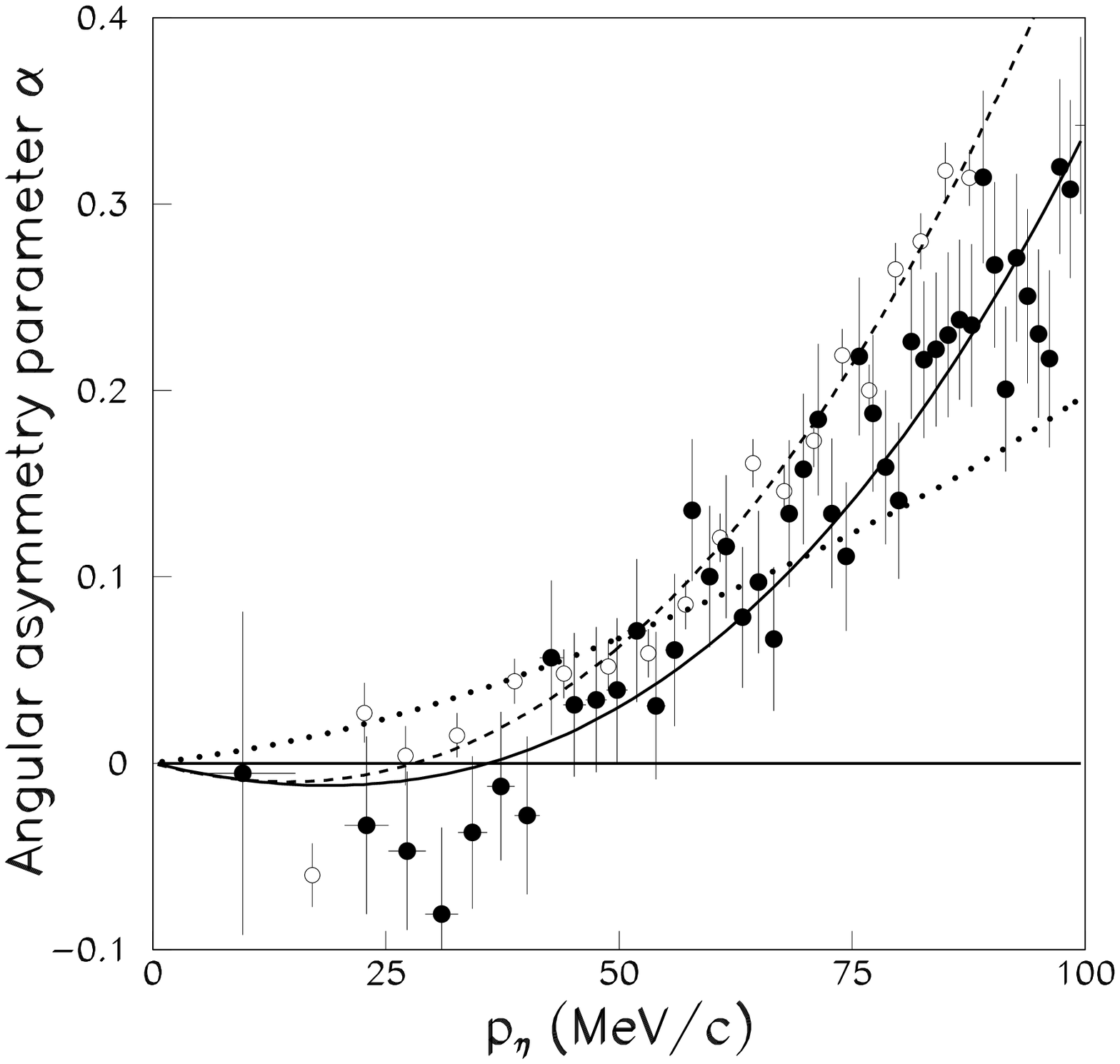}} 
        \caption{(left) Total cross section for the $dp\to ^{3}\!\!{\mathrm{He}}\eta$ reaction 
as determined by the
COSY-ANKE~\cite{timo} (open circles) and the COSY-11~\cite{jurek-he3} 
(full dots) and~\cite{adam} (triangles). The solid and dashed lines represent the scattering length fit 
to the COSY-11 and COSY-ANKE data, respectively.
(right) Angular asymmetry parameter $\alpha$. Closed and open circles represent results of
COSY-ANKE~\cite{timo} and COSY-11~\cite{jurek-he3}, respectively.
The dashed and solid lines denote results~\cite{wilkin2} 
of the fit (allowing for the phase variation) to the COSY-11 and COSY-ANKE data, respectively.
The dotted line denotes result of the fit  without the  phase variation.
The figure is adapted from Ref.~\cite{wilkin2}.} 
\label{fig1} 
    \end{center} 
\end{figure}

In 1985 Bhalerao and Liu~\cite{liu1} performed a coupled-channel 
analysis of the $\pi {\mathrm{N}}\to \pi {\mathrm{N}}$, 
$\pi {\mathrm{N}}\to \pi \pi {\mathrm{N}}$ and $\pi {\mathrm{N}}\to \eta {\mathrm{N}}$ reactions 
and came to conclusion that the interaction between the nucleon and the $\eta$ meson
 is attractive. 
Based on this finding Haider and Liu  postulated the existence of the 
$\eta$--mesic nuclei~\cite{liu2}, in which the electrically neutral $\eta$ meson 
might be bound with the nucleons by the strong interaction. 
The formation of such a bound state can only take place if
the real part of the $\eta$-nucleus scattering length is negative 
(attraction), and the 
magnitude of the real part  is greater than the 
magnitude of the imaginary part~\cite{liu3,sibirtsev}. 
In the 1980's  
the $\eta$--mesic nuclei were considered
to exists for A~$\ge$~12~only~\cite{liu2} 
due to the  
relatively small value of the $\eta$N scattering length
($a_{\eta {\mathrm{N}}}~=~(0.28\, +\, i0.19)$~fm~\cite{liu1}).
However, recent theoretical investigations of hadronic- and photo-production 
of the $\eta$ meson result in values of the real part of $a_{\eta {\mathrm{N}}}$ 
which depending on the
analysis method range from
0.25~fm up 
to 1.05~fm~\cite{wycech}, 
and which do not exclude the formation 
of bound $\eta$-nucleus states  for such light nuclei as helium~\cite{wilkin1,wilkinsymposium,wycech1} 
or even for deuteron~\cite{green}.
According to the calculations including multiple scattering theory~\cite{wycech1}
or Skyrme model~\cite{scoccola} an especially good candidate for binding 
is the $^4{\mathrm{He}} -\eta$ system.
Recent calculations by Haider~\cite{haideracta} or Tryasuchev and Isaev~\cite{tryasuchev} 
also indicate the binding in the $^4{\mathrm{He}} -\eta$ system,
while they rather exclude the existence of the $^3{\mathrm{He}} - \eta$ state. 
On the other hand there are promising experimental signals 
which may be interpreted as indications of the the $^3{\mathrm{He}} -\eta$ bound state,
as for example the shape of the excitation function 
for the $d\,p\to ^3\!\!{\mathrm{He}}\,\eta$ reaction~\cite{wilkin1}
determined by the SPES-4~\cite{berger}, SPES-2~\cite{mayer}, 
COSY-11~\cite{jurek-he3}, and COSY-ANKE~\cite{timo} collaborations~(Fig.~\ref{fig1}(left)). 
It has been indicated by Wilkin~\cite{wilkin2,wilkinaip} that a steep rise of 
the total cross section in the very close-to-threshold region followed 
by a plateau may be due to the existence of a pole 
of the $\eta ^3{\mathrm{He}}\to \eta ^3{\mathrm{He}}$ 
scattering amplitude in the complex excess energy 
plane $Q$ with $Im(Q)<0$~\cite{wilkin2}. This reference shows that the 
occurrence of the pole changes the phase and the magnitude of the s-wave production amplitude. 
And indeed, the momentum dependence of the asymmetry in the angular distributions 
of $\cos\theta_{\eta}$ expressed in terms of the asymmetry
parameter $\alpha$ ~\cite{jurek-he3}, 
can only be satisfactorily described (solid and dashed lines in Fig.~\ref{fig1} (right)) 
if a very strong phase variation associated with the pole is included 
in the fits~\cite{wilkin2,wilkinaip}. 
Otherwise there is a significant discrepancy between the experimental data and 
the theoretical description (dotted line in Fig.~\ref{fig1} (right)).

\section{Study of the $\eta-{^3{\mathrm{He}}}$ system in $d-p$ collisions at COSY-11}
Details of the experimental technique~\cite{brauksiepe,AIP,hab} and a comprehensive description 
of results concerning the search of the $\eta$-mesic $^3{\mathrm{He}}$ nucleus  
conducted by the COSY-11 group 
were described elsewhere~\cite{c11acta,c11meson08,c11nuclphys,c11aps}.
Therefore, here we only summarize the main results.
The measurements were carried out
using the deuteron beam of COSY which was circulating through 
the stream of the internal hydrogen target
of the cluster-jet type~\cite{brauksiepe}.
Data were taken during a slow  acceleration of the beam
from 3.095~GeV/c to 3.180~GeV/c, crossing the kinematical threshold
for the $\eta$ production in the $dp \rightarrow {^3{\mathrm{{He}}}}\,\eta$ reaction
at 3.141~GeV/c. 
The corresponding variation of the excess energy in the $\eta-{^3{\mathrm{{He}}}}$ 
system ranged from -10~MeV to +9~MeV.
For the scanned beam momentum interval we determined excitation functions 
for the pion production in the $dp \rightarrow {^3\mbox{He}}\, \pi^0$ 
and $dp\rightarrow ppp\pi^-$ reaction which were chosen out of 
possible decay channels of the $\eta$-mesic ${^3{\mathrm{He}}}$.
A signature of existence of the $\eta-{^3{\mathrm{He}}}$ bound state
was based on observation of a resonance like structure with the center lying below
the $\eta$ production threshold in the measured excitation curves.
 
For the $dp \rightarrow {^3\mbox{He}}\, \pi^0$ channel we concentrated on 
differential cross sections for the forward pion angles  
($\Theta_{d-\pi}^{cm} = 0^{\circ}$). 
This choice was dictated by the fact that the $dp \rightarrow {^3\mbox{He}} \pi^0$
cross section is up to two orders of magnitude smaller at the forward angles
than at the most backward angles.
Assuming that the searched structure is produced isotropically,
one can expect that it can be best seen just at the forward angles since
it appears on the level of small ``non-resonant'' cross section.
The determined excitation function for the  $dp \rightarrow {^3\mbox{He}}\,\pi^0$
process does not show any structure
which could originate from a decay of $\eta-{^3\mbox{He}}$ bound
state \cite{c11nuclphys}. 
The estimated upper limit for the cross section of the 
$dp \to (\eta^3\mbox{He})_{bound} \to {^3\mbox{He}}\,\pi^0$ 
reaction chain is equal to 70~nb. 
This limit appears not very restrictive at least under assumption
that the cross sections for the ${^3\mbox{He}}-\eta$ bound state formation are of the same
order as the $dp \rightarrow {^3\mbox{He}}\, \eta$ cross sections near threshold (0.4~$\mu$b), 
and that other possible decay channels like  $dp \pi^0$ or $ppp\pi^-$are more favorable.

\begin{figure}[h] 
    \begin{center} 
        {\includegraphics[scale=0.4]{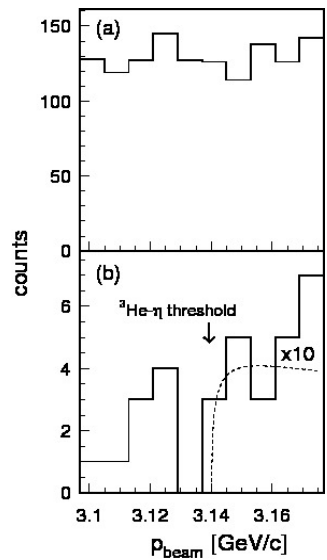}}
        \caption{Number of $dp\rightarrow ppp\pi^-$ events as a function
of the beam momentum without any cuts (a) and after rejection of events
corresponding to the quasi-free $\pi^-$ production (b). 
The dashed line represents calculations
of the $\eta$ absorption described in the text.} 
\label{ppppi_counts} 
    \end{center} 
\end{figure}

We expect that $ppp \pi^-$ is one of the favorable decay channels
of the $\eta-{^3\mbox{He}}$ bound state since it corresponds to one step process
of absorption of the $\eta$ meson on the neutron
inside the $^3\mbox{He}$ nucleus in the reaction
$\eta n \rightarrow N^*(1535) \rightarrow p \pi^-$.
In the $N^*$ rest frame the pion and the proton are emitted back-to-back with momenta of about 430~MeV/c.
In the center-of-mass (c.m.) system
these momenta are smeared due to the Fermi motion of the neutron
inside the $^3\mbox{He}$ nucleus.
However, they are significantly larger than the momenta of the two remaining protons
which are in the order of 100~MeV/c,
and  which play the role of "spectators".
The counting rate of all identified $dp\rightarrow ppp\pi^-$ events including the
quasi-free $\pi^-$ production stays constant in the scanned range 
of the beam momentum (see Fig.~\ref{ppppi_counts}a). 
However, after rejection of the quasi-free events,
the number of $dp\rightarrow ppp\pi^-$ counts in the beam momentum interval above
the $\eta$ threshold is higher than the number of counts in the interval
of equal width below the threshold (see Fig.~\ref{ppppi_counts}b).
This difference is equal to $23-9=14$ and its statistical significance 
is of $2.5 \sigma$. The observation of this effect was reported in Ref.~\cite{c11meson08}.
As a possible reaction mechanism explaining the observed enhancement 
we suggested the production of "on-shell" $\eta$ mesons 
in the reaction $dp \rightarrow {^3\mbox{He}}\,\eta$
which subsequently are absorbed in the $^3\mbox{He}$ nucleus and converted to 
negatively charged pions in the interaction with the neutron.

For testing the $\eta$ absorption hypothesis
we estimated the counting rate of the $dp\rightarrow ppp\pi^-$ events
assuming that the absorption cross section $\sigma_{abs}$
is equal to the cross section for the $dp \rightarrow {^3\mbox{He}}\,\eta$ reaction
(for more details see Ref.\ \cite{c11acta}).
This assumption is justified by the observation that the real and imaginary part
of the ${^3\mbox{He}}-\eta$ scattering length \cite{jurek-he3} have comparable values.
The result of our estimation multiplied by a factor of 10 is shown by the dashed line 
in Fig.~\ref{ppppi_counts}b.
It underestimates the experimental counts roughly by an order of magnitude
and thus it does not corroborate  the $\eta$ absorption hypothesis.

For testing this hypothesis
we analysed also  momentum and angular distributions of the final state
particles in the $dp \rightarrow ppp \pi^-$ reaction.
The experimental distribution of the pion momentum in the c.m. system 
is centred around 430~MeV/c  
as expected for pions originating from the decay of the $N^*(1535)$.
It agrees with results of simulations of the $\eta$ absorption.
However, the momentum distribution of the proton 
with the highest momentum, which we call the leading proton, 
is centred at around 300~MeV/c and is lower
than the corresponding distribution for the $\eta$ absorption grouped around 430~MeV/c.
A disagreement takes place also in the case of
the distribution of the center-of-mass angles between the pion momentum vector
and the momentum vector of the leading proton 
which in the case of simulations of the $\eta$ absorption 
are close to $180^{\circ}$ and for the experiment lie around $160^{\circ}$.

One can expect, that due to very similar kinematical conditions,
absorption of the $\eta$ mesons bound in the $^3\mbox{He}$ nucleus
is characterized by differential distributions 
which are very close to ones predicted in our simulations
for the absorption of "on-shell" $\eta$ mesons.
In particular one can expect that the $\pi^--p$ pairs originating
from the decay of the $\eta$-mesic $^3\mbox{He}$ are emitted at c.m. angles
concentrated predominately in the range $150^{\circ}-180^{\circ}$ 
as it is the case for the $\eta$ meson absorption. 
In this angular range, there are only two experimental counts.
Assuming, that these two counts originate from the decay of the $^3\mbox{He}-\eta$ bound state,
we estimated the cross section for the production of such a state in $d-p$ collisions
close to the $\eta$ production threshold. 
The resulting cross section of  $0.27\pm0.19$~$\mu$b 
should be considered as an upper limit for the production cross section 
of the $\eta$-mesic ${^3\mbox{He}}$ since the observed two events might 
originate from other processes than the bound state decay.

Data collected in the COSY-11 measurements with the ramped deuteron beam 
were also used to investigate the cusp effect observed at SATURNE 
in the threshold excitation curve for the process 
$dp \rightarrow {^3\mbox{He}}\, X$ \cite{c11saturne}.
As suggested by Wilkin \cite{c11wilk}, the cusp visible in the SATURNE data 
below the $\eta$ threshold could be caused by an interference between 
an intermediate state including the $\eta$ meson and the non-resonant background 
corresponding to the multi-pion production.
In the SATURNE experiment the ${^3\mbox{He}}$ ejectiles were measured with the SPES~IV spectrometer
which was set in such a way that it registered the ${^3\mbox{He}}$ which was approximately at rest 
in the c.m. frame. 
Since the COSY-11 momentum and angular acceptance was much larger than one of the SPES~IV spectrometer,
the limitation on the c.m. momenta of ${^3\mbox{He}}$ was realized by means of corresponding cuts 
during the data analysis \cite{c11aps}.
Contrary to the SATURNE result no cusp below the $\eta$ threshold was observed \cite{c11nuclphys}.

\section{Search of the $\eta-{^4{\mathrm{He}}}$ state with WASA-at-COSY}
\begin{figure}[h]
{\includegraphics[scale=0.300]{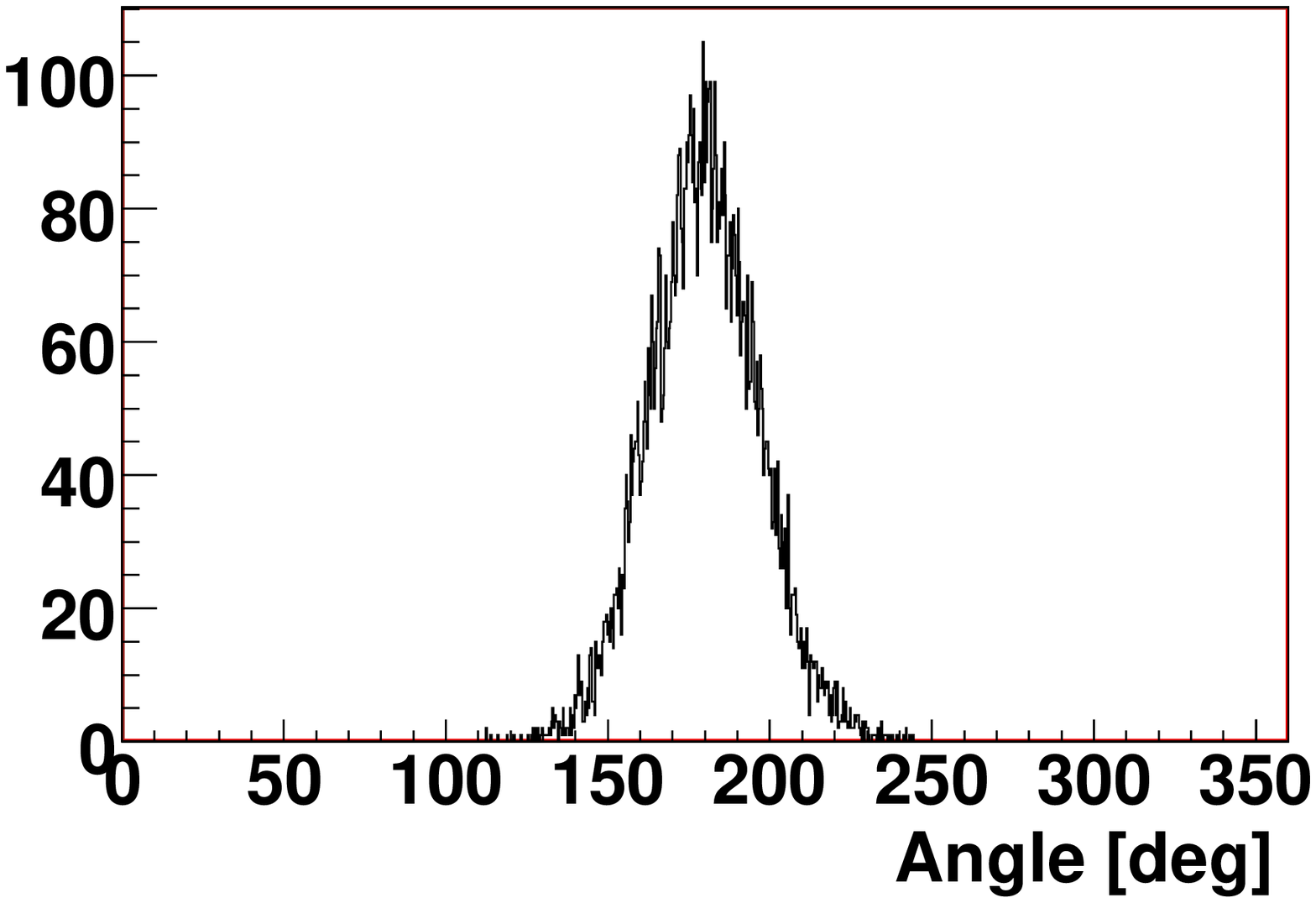}} 
{\includegraphics[scale=0.300]{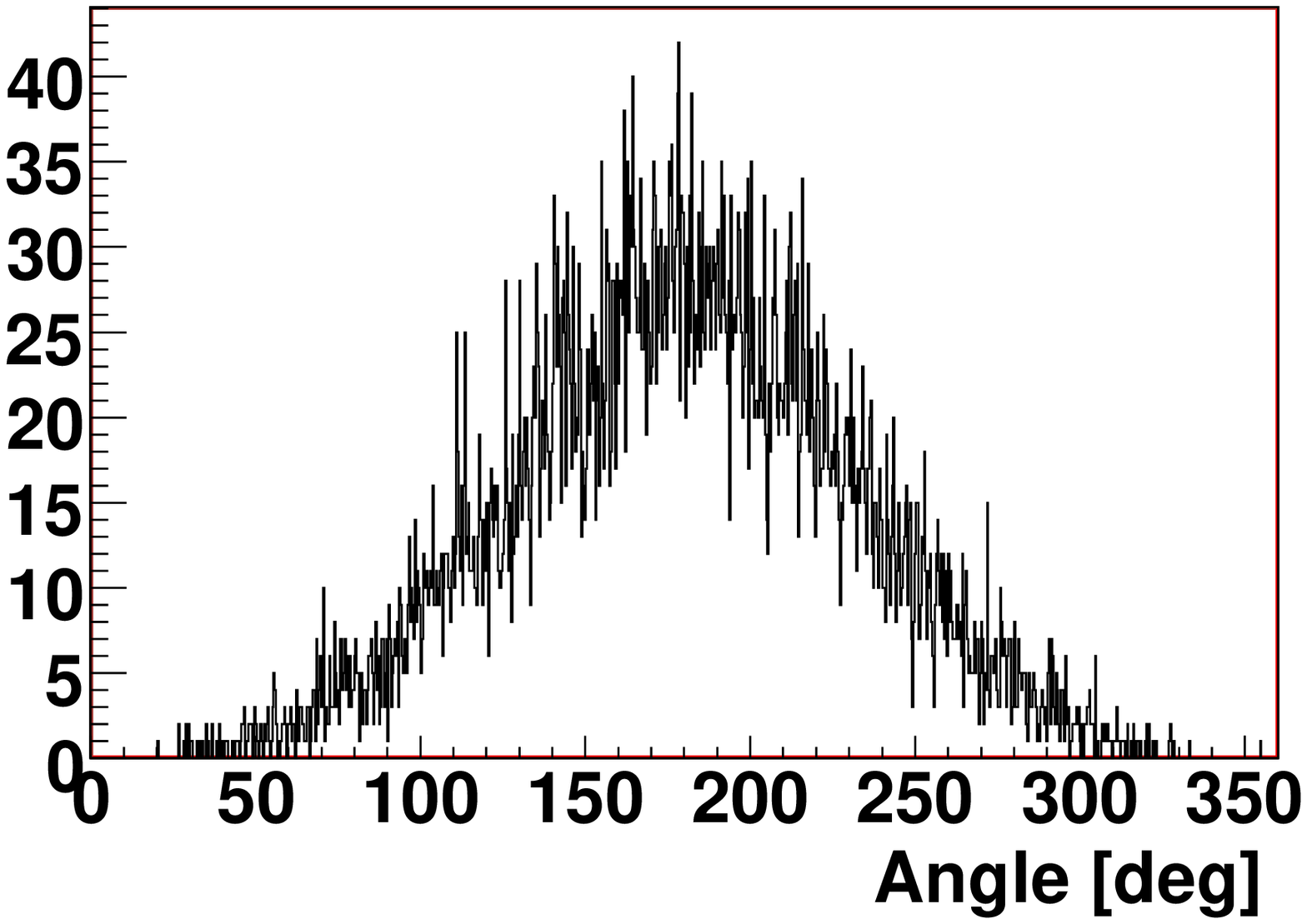}}
\caption{\label{angle} 
Distribution of the relative $p-\pi$  angle seen in the reaction c.m. system
as simulated for the processes leading to the creation of the eta-helium bound state: 
 $dd\to(^4He\eta)_{bound}\to^3\!\!Hep\pi$ (left),
and for the prompt production of the $^3\!Hep\pi$ system assuming a homogeneous
population of the phase space for the  $dd\to^3\!\!Hep\pi$ reaction (right).
The figure is adapted from reference~\cite{wasaacta}.}
\end{figure}

The installation of the WASA detector at COSY opened a unique possibility to search for
the $^4{\mbox{He}}-\eta$ bound state with  high statistics and high acceptance.
We  conduct a search 
via an exclusive measurement
of the excitation function for the $dd \rightarrow {^3\mbox{He}}\, p\, \pi^-$ reaction
varying  continuously
the beam momentum
around the 
threshold for the $d\,d \to ^4\!{\mathrm{He}}\,\eta$ reaction.
Ramping of the beam momentum and 
taking advantage of the large
acceptance of the WASA detector\footnote{For the coincidence registration of all ejectiles
from the $dd \to (\eta\,{^4\mbox{He}})_{bound} \to {^3\mbox{He}}\, p\, \pi^-$ 
reaction the acceptance of the WASA-at-COSY detector equals to almost 70\%.} 
allows to minimize systematical uncertainties making
the WASA-at-COSY a unique facility~\cite{wasa1}  for such kind of exclusive experiments.
The  ${^4\mbox{He}}-\eta$ 
bound state should manifest itself as a resonant like structure
below the threshold for the $dd \to {^4\mbox{He}}\,\eta$ reaction.
If a peak below the ${^4\mbox{He}}\,\eta$ threshold
is found, then
the profile of the 
excitation curve  will allow 
to determine the binding energy and the width 
of the $^4{\mathrm{He}}-\eta$ bound state.
If, however, only an enhancement around the threshold is found, then  it will enable
to establish the relation between width and binding energy~\cite{osetPLB550}.
Finally, if no structure is seen
the upper limit for 
the cross section of the production 
of the $\eta$-helium nucleus 
will be set. 
In addition, when searching for the signal of the $\eta$-mesic state
we may take advantage of the fact that 
the distribution of the relative angle between the $nucleon-pion$ pair
for the background 
(due to the prompt $dd\to {^3{\mbox{He}}}\, p\, \pi^-$ reaction) is much broader 
than the one expected from the decay of the bound state.
This is because the relative angle between the outgoing $nucleon-pion$ pair  originating 
from the decay of the N$^*(1535)$ resonance 
is equal to 180$^{\circ}$ in the N$^*$ reference frame and it is 
smeared only by about 30$^{\circ}$ in the reaction center-of-mass frame 
due to the Fermi motion of the nucleons inside the $\mbox{He}$ nucleus.
Fig.~\ref{angle} shows the distribution of the relative $proton-pion$ angle 
as expected for the signal and for the background due to the prompt 
$dd\to {^3{\mbox{He}}}\, p\, \pi^-$ reaction.

\begin{figure}[h] 
    \begin{center} 
        {\includegraphics[scale=0.300]{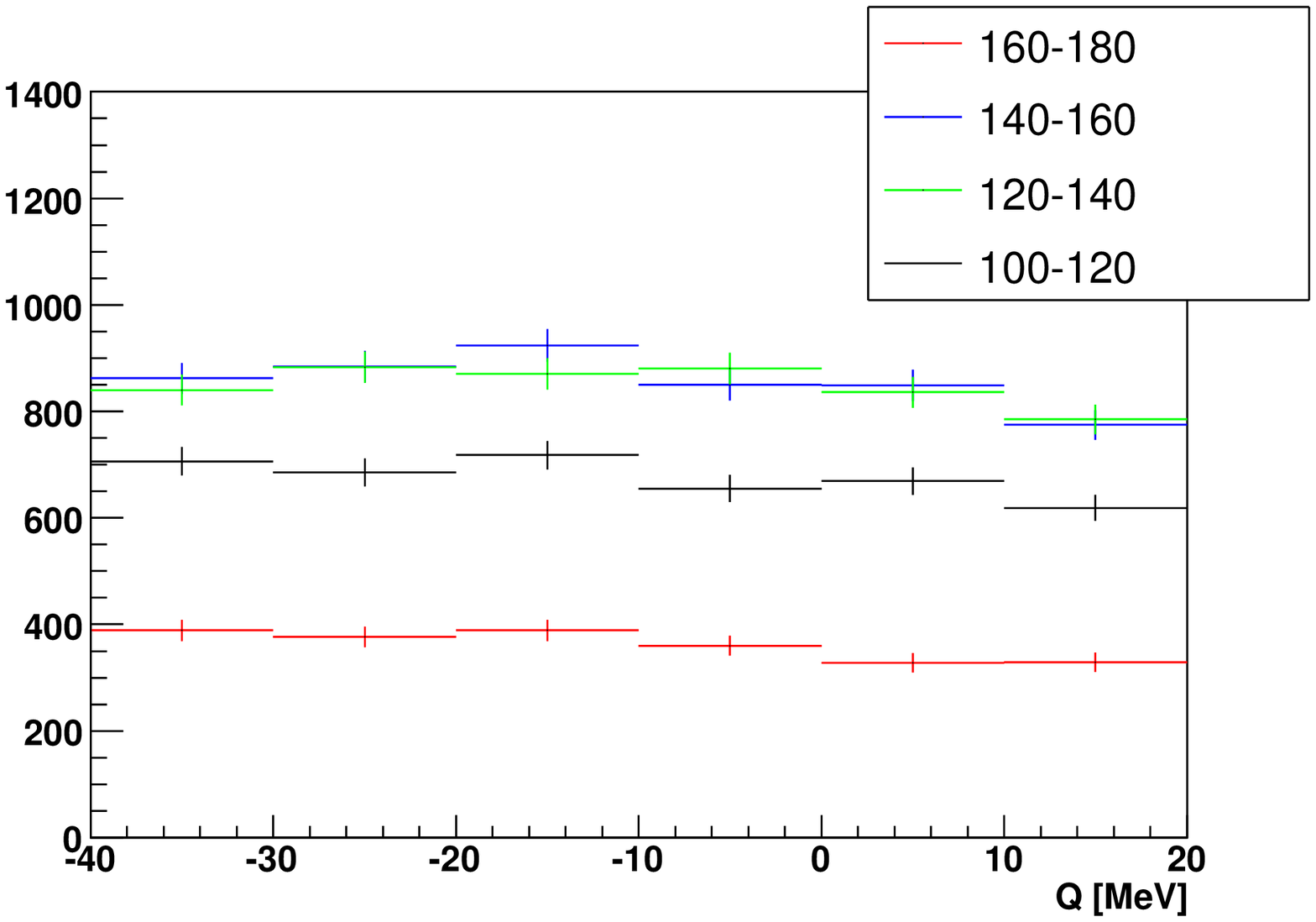}} \hfill
        {\includegraphics[scale=0.303]{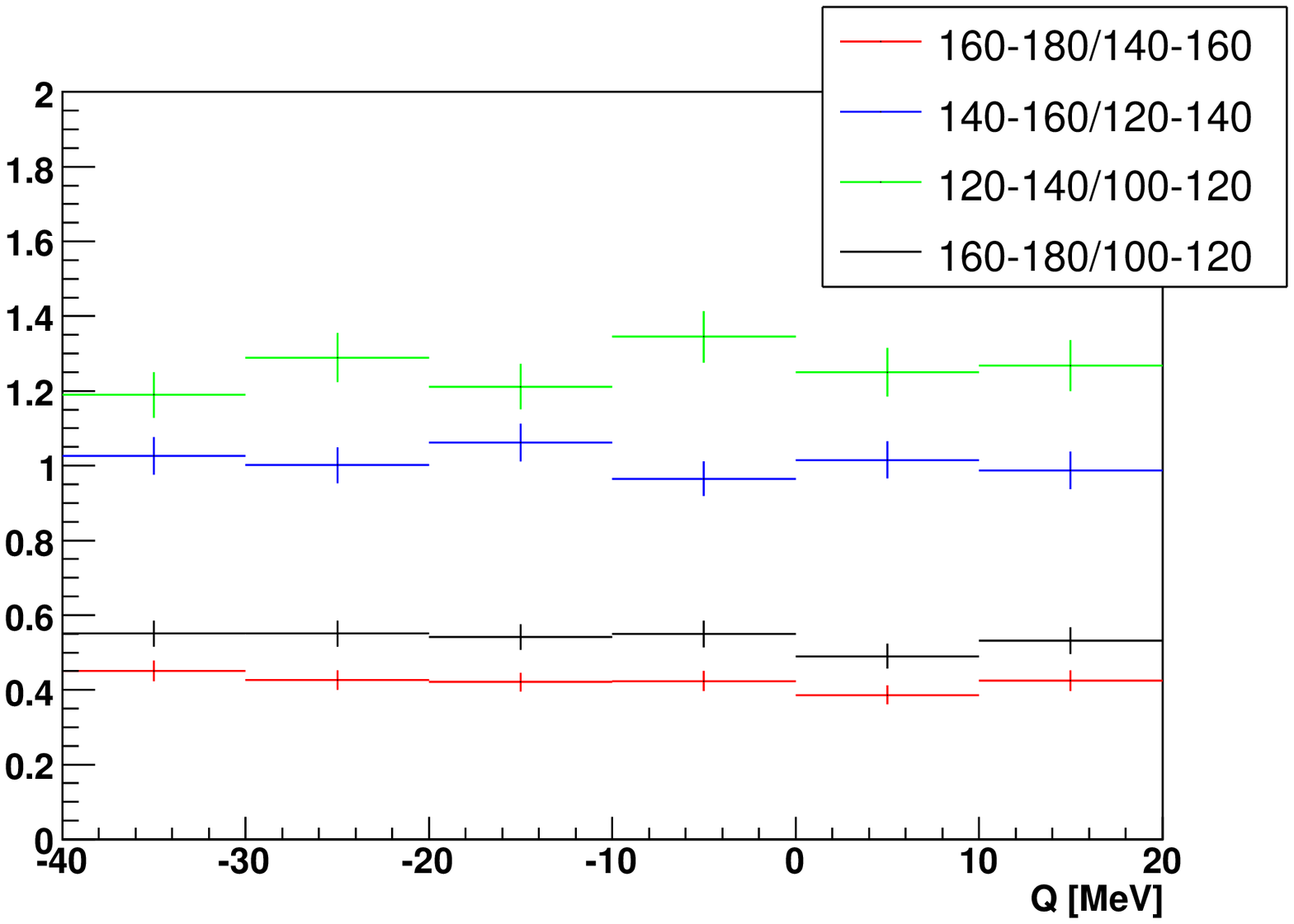}} 
        \caption{(left) Excitation function for 
the  $dd\to^3\!\!Hep\pi$ reaction measured in the 20 degrees intervals of the 
 $\Theta_{cm}(p-\pi)$ angle. 
Courtesy to~\cite{wojciech}. (Right) Ratio of excitation functions angular ranges as indicated in the figure. 
\label{fig4}} 
    \end{center} 
\end{figure}

In the first experiment conducted in June 2008, 
we used a deuteron pellet target and the COSY deuteron 
beam with a ramped momentum corresponding to a variation of the excess energy 
for the $^4{\mathrm{He}}-\eta$ system from -51.4~MeV to 22~MeV.
At present the data are evaluated and preliminary 
excitation curves determined for few  intervals of the 
 $\Theta_{cm}(p-\pi)$ angle are shown in the left panel of Fig.~\ref{fig4}.
The figure indicates no structure in the angular range close to the 180 degree where 
the signal is expected. The ratio of excitation functions from various angular ranges
is also constant as indicated in the right panel of Fig.~\ref{fig4}. 
Therefore, taking into account the luminosity and the detector acceptance we preliminary estimated 
that an upper limit for the eta-mesic helium production via 
the $dd \to (\eta\,{^4\mbox{He}})_{bound} \to {^3\mbox{He}}\, p\, \pi^-$ reaction is equal to  about 
20~nb on a one sigma level.
The experiment will be continued in November 2010. However, we cannot  estimate the 
statistics required for the observation of the signal since it is  
 a priori impossible to predict in a model independent way a cross section 
for the creation of the eta-mesic nucleus, which strongly depends on the strength 
of the relatively poorly known $\eta$-nucleon interaction, 
and which involves computations of the behaviour 
of the many-body system. 
Therefore, based on the fact that the real and imaginary part of the helium-$\eta$
scattering length are almost of the same value 
we estimate the cross section for the creation of the eta-mesic helium applying
the assumption that the probability of the production 
of the $\eta$ meson in continuum
and its  absorption on the helium nucleus is of the same order.
This leads to the hypothesis that the cross section for the creation
of the bound state below the threshold (which is than connected with the
absorption of the $\eta$ meson on the helium nucleus)
is in the first order the
same as the close to threshold cross section for the $\eta$ meson production.
Therefore,  for the  estimation of the counting rate in the future measurements
we assume that the cross section in the maximum of the Breit-Wigner distribution
for the  $dd \rightarrow ({^4\!He}-\eta)_{bound}$ production is about 15 nb
as measured for the $dd\to ^4\!He\eta$ reaction~\cite{Frascaria94,Willis97,wronskaacta}.
Further on, we assume, that the probability of decay of the ${^4\mbox{He}}-\eta$ bound state
into the ${^3\mbox{He}} p \pi^-$ channel is equal to $\frac{1}{4} \cdot \frac{1}{2}=\frac{1}{8}$.
The factor $\frac{1}{4}$ takes into account the fact that there are four possible
$\eta$ absorption channels:
$\eta p \rightarrow p \pi^0$,
$\eta p \rightarrow n \pi^+$,
$\eta n \rightarrow n \pi^0$,
$\eta n \rightarrow p \pi^-$.
In turns, the factor $\frac{1}{2}$ represents our guess of the probability
that the three observer nucleons ($ppn$), in the process
of the $\eta$ absorption on the neutron in ${^4\mbox{He}}$,
form  ${^3\mbox{He}}$ in the final state.
Estimation of this effect needs, in our understanding of the problem, projection
of the ${^4\mbox{He}}$ wave function on the
wave function of the ${^3\mbox{He}}-n$ pair and inclusion of the pion and proton rescattering
on the observer nucleons.
Since, this kind of estimations isn't at present in our reach,
we justify our guess using an analogy between
the decay of the ${^4\mbox{He}}-\eta$ nuclei
and the ${^4_{\Lambda}\mbox{He}} $-hypernuclei.
For the later case it was observed namely that in the $\pi^-$ decay channel
the decay mode ${^4_{\Lambda}\mbox{He}} \rightarrow \pi^- p {^3\mbox{He}}$
is dominant \cite{Fet}.
The above considerations implies that the cross section for the
$dd \to (^4\!He\eta)_{bound} \to ^3\!\!He p \pi^-$ reaction should 
be in the order of 2~nb.
This level of  sensitivity can be reached in the  experiment which will be continued in
November 2010~\cite{wasaproposal}.

\section{Search of the bound state with quasi-free beams}
The systematic uncertainties in establishing the shape of the excitation functions
discussed in the previous sections are significantly reduced 
by using the momentum ramping technique since
the energy range of interest for the search of the $\eta$-mesic nucleus is scanned in each COSY cycle.

\vspace{0.0cm}
\begin{figure}[h]
\centering
\includegraphics[width=9.5cm,height=3.3cm]{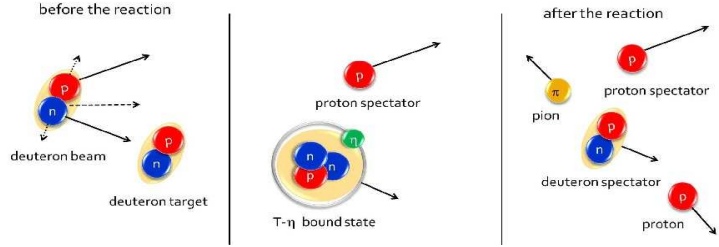}
\caption{Schematic picture of the quasi-free 
$nd\rightarrow(\mbox{T}$-$\eta)_{bs}\rightarrow d p \pi{}^{-}$ reaction. 
The Fermi momentum of the nucleons inside the deuteron 
is  presented by the dotted arrows and the beam momentum by the dashed one.
The figure is adapted from reference~\cite{magdalena}.
\label{fig2}}
\end{figure}
A~scan of the energy can also be achieved in the case of the fixed beam energy
available for the external beam experiments. This may be realised by means 
of the quasi-free reactions as already successfully used at COSY for the study
of the meson production in the quasi free proton-neutron collisions~\cite{prc2009,tof}.
For instance, a search of the  $\eta$-mesic tritium may be realized 
by studying the excitation 
function of the $n\, d \to (\eta\, T)_{bound} \to d\, p\, \pi^-$   reaction
using a deuteron beam and tagging the $nd$ reactions
by the measurement of the spectator protons ($p_{sp}$) from the 
$d\, d\to p_{sp}\, n\, d\to p_{sp}\, (\eta\, T)_{bound} \to p_{sp}\, d\, p\, \pi^-$ reaction.
This reaction, shown schematically in Fig.~\ref{fig2}, may be measured using the COSY-TOF detector.
The Fermi motion of nucleons inside the deuteron beam will allow to scan a large 
range (in the order of 100~MeV) of excess energies in the $nd$ reaction.

\section{Acknowledgements}
The work was partially supported by the
European Co\-mmu\-nity-Research Infrastructure Activity
under the FP6 and FP7 programmes (Hadron Physics,
RII3-CT-2004-506078, PrimeNet No. 227431.), by
the Polish Ministry of Science and Higher Education under grants
No. 3240/H03/2006/31  and 1202/DFG/2007/03,
by the German Research Foundation (DFG),
by the FFE grants from the Research Center J{\"u}lich,
and by the virtual institute "Spin and strong QCD" (VH-VI-231).

\end{document}